\documentclass[preprint,superscriptaddress]{revtex4-2}
\usepackage[T1]{fontenc}
\usepackage{times}
\usepackage{graphicx}
\usepackage{booktabs}

\usepackage{amsfonts, amsmath,amssymb, bm, graphicx}
\usepackage{siunitx}

\usepackage{amssymb}
\usepackage[english]{babel}
\usepackage{lipsum}

\newcommand{\figref}[2]{\hyperref[#1]{\ref{#1}(#2)}}
\newcommand{\figrefsub}[3]{\hyperref[#1]{\ref{#1}(#2)#3}}

\usepackage{physics}
\usepackage{lipsum}
\usepackage{changes}

\usepackage[colorlinks=true,allcolors=blue]{hyperref}
\usepackage{footmisc}

\usepackage{upgreek}

\usepackage{soul}

\makeatletter
\let\ORIbbl@fixname\bbl@fixname
\def\bbl@fixname#1{%
  \@ifundefined{languagealias@\expandafter\string#1}
    {\ORIbbl@fixname#1}
    {\edef\languagename{\@nameuse{languagealias@#1}}}%
}
\newcommand{\definelanguagealias}[2]{%
  \@namedef{languagealias@#1}{#2}%
}
\makeatother

\definelanguagealias{en}{english}

\begin{document}

\title{Pattern recognition with a magnon-scattering reservoir}

\author{Lukas K\"orber}
\affiliation{Helmholtz-Zentrum Dresden--Rossendorf, Institut f\"ur Ionenstrahlphysik und Materialforschung, D-01328 Dresden, Germany}
\affiliation{Fakultät Physik, Technische Universität Dresden, D-01062 Dresden, Germany}

\author{Christopher Heins}
\affiliation{Helmholtz-Zentrum Dresden--Rossendorf, Institut f\"ur Ionenstrahlphysik und Materialforschung, D-01328 Dresden, Germany}
\affiliation{Fakultät Physik, Technische Universität Dresden, D-01062 Dresden, Germany}

\author{Tobias Hula}
\affiliation{Helmholtz-Zentrum Dresden--Rossendorf, Institut f\"ur Ionenstrahlphysik und Materialforschung, D-01328 Dresden, Germany}
\affiliation{Institut für Physik, Technische Universität Chemnitz, 09107 Chemnitz}

\author{Joo-Von Kim}
\affiliation{Centre de Nanosciences et de Nanotechnologies (CNRS), Université Paris-Saclay, Palaiseau, 91120,France}

\author{Helmut Schultheiss}
\affiliation{Helmholtz-Zentrum Dresden--Rossendorf, Institut f\"ur Ionenstrahlphysik und Materialforschung, D-01328 Dresden, Germany}
\affiliation{Fakultät Physik, Technische Universität Dresden, D-01062 Dresden, Germany}

\author{Jürgen Fassbender}
\affiliation{Helmholtz-Zentrum Dresden--Rossendorf, Institut f\"ur Ionenstrahlphysik und Materialforschung, D-01328 Dresden, Germany}
\affiliation{Fakultät Physik, Technische Universität Dresden, D-01062 Dresden, Germany}

\author{Katrin Schultheiss}
\affiliation{Helmholtz-Zentrum Dresden--Rossendorf, Institut f\"ur Ionenstrahlphysik und Materialforschung, D-01328 Dresden, Germany}

%\author{R. Verba}
%\affiliation{Institute of Magnetism, National Academy of Sciences of Ukraine, Kyiv 03142, Ukraine}

%\author{T. Hache}
%\affiliation{Helmholtz-Zentrum Dresden--Rossendorf, Institut f\"ur Ionenstrahlphysik und Materialforschung, D-01328 Dresden, Germany}
%\affiliation{TU Chemnitz, Germany}

%\author{L. Bischoff}
%\affiliation{Helmholtz-Zentrum Dresden--Rossendorf, Institut f\"ur Ionenstrahlphysik und Materialforschung, D-01328 Dresden, Germany}

%\author{A. Awad}
%\affiliation{Department of Physics, University of Gothenburg, 412 96 Gothenburg, Sweden}

%\author{V. Tiberkevich}
%\affiliation{Department of Physics, Oakland University, Rochester, MI 48309, USA}

%\author{A.N. Slavin}
%\affiliation{Department of Physics, Oakland University, Rochester, MI 48309, USA}

%\author{J. Fassbender}
%\affiliation{Helmholtz-Zentrum Dresden--Rossendorf, Institut f\"ur Ionenstrahlphysik und Materialforschung, D-01328 Dresden, Germany}
%\affiliation{TU Dresden, D-01062 Dresden, Germany}

\date{\today}

\begin{abstract}

Magnons are elementary excitations in magnetic materials and undergo nonlinear multimode scattering processes at large input powers. In experiments and simulations, we show that the interaction between magnon modes of a confined magnetic vortex can be harnessed for pattern recognition. We study the magnetic response to signals comprising sine wave pulses with frequencies corresponding to radial mode excitations. Three-magnon scattering results in the excitation of different azimuthal modes, whose amplitudes depend strongly on the input sequences. We show that recognition rates above 95\% can be attained for four-symbol sequences using the scattered modes, with strong performance maintained with the presence of amplitude noise in the inputs.
\end{abstract}

\maketitle
A key challenge in modern electronics is to develop low-power solutions for information processing tasks such as pattern recognition on noisy or incomplete data. One promising approach is physical reservoir computing, which exploits the nonlinearity and recurrence of dynamical systems (the `reservoir') as a computational resource~\cite{Maass:2002kf}.  Examples include a diverse range of materials such as water~\cite{Fernando:2003ft}, optoelectronic systems~\cite{Appeltant:2011jy, Paquot:2012dx, Ortin:2015au, Larger:2017cx}, silicon photonics~\cite{Vandoorne:2014hv}, microcavity lasers~\cite{Sunada:2019pr}, organic electrochemical transistors~\cite{Cucchi:2021rc}, dynamic memristors~\cite{Zong:2021dm}, nanowire networks~\cite{Milano:2022im}, and magnetic devices~\cite{Torrejon:2017hj, Nakane:2018jc, Taro:2019cp, Watt:2021ia, Nakane:2021sw, Gartside:2022rt, ababei2021neuromorphic}.

\begin{figure*}[t!]
    \centering
    \includegraphics[width=\textwidth]{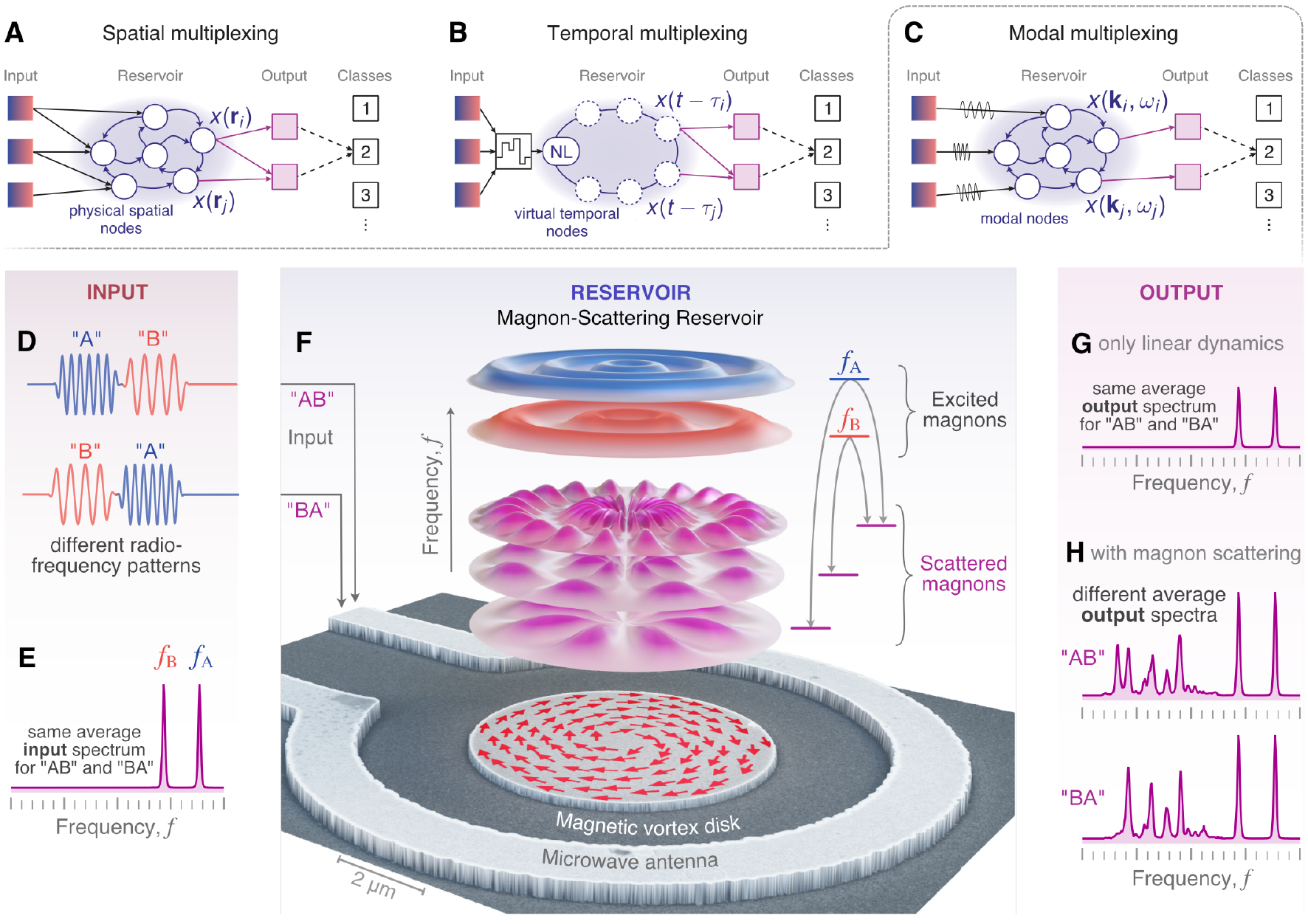}
    \caption{\textbf{Working principle of a magnon-scattering reservoir (MSR).} Sketches of different reservoirs based on (\textbf{A}) spatial, (\textbf{B}) temporal and (\textbf{C}) modal multiplexing, the concept behind the MSR. (\textbf{D}) Radiofrequency pulses with different temporal order but (\textbf{E}) the same average frequency content  are used to trigger (\textbf{F}) nonlinear scattering between the magnon eigenmodes in a magnetic vortex disk. The dynamic response is experimentally detected using Brillouin-light-scattering microscopy (see supplementary materials). In contrast to  a linear system (\textbf{G}),  the MSR produces different outputs depending on the temporal order of the input (\textbf{H}).}
    \label{fig:FIG1}
\end{figure*}

% Explain what a reservoir is
The physical reservoir embodies a recurrent neural network. A natural implementation comprises interconnected nonlinear elements in space (spatial multiplexing, Fig.~\ref{fig:FIG1}A), where information is fed into the system via input nodes representing distinct spatial elements, and the dynamical state is read out through another set of output nodes\cite{Fernando:2003ft, Cucchi:2021rc, Kan:2021sr, Milano:2022im}. Another approach involves mapping the network onto a set of virtual nodes in time by using delayed-feedback dynamics on a single nonlinear node (temporal multiplexing, Fig.~\ref{fig:FIG1}B)~\cite{Appeltant:2011jy, Paquot:2012dx, Ortin:2015au, Larger:2017cx, Zong:2021dm, Watt:2021ia}, which reduces the complexity in spatial connectivity at the expense of more intricate time-dependent signal processing.

% What is different here
Here, we study an alternative paradigm in which we exploit instead the dynamics in the \textit{modal space} of a magnetic element. This scheme relies on magnon interactions in magnetic materials whereby inputs and outputs correspond to particular eigenmodes of a micromagnetic state. Processes such as three-magnon scattering interconnect the modes with each other and, with that,  provide the nonlinearity and recurrence required for computing. We refer to this approach as modal multiplexing with signals evolving in reciprocal space, in which the actual computation is performed. This is distinct from other wave-based schemes where information is processed with wave propagation and interference in real space~\cite{Nakane:2018jc, Sunada:2019pr, Marcucci:2020to, Nakane:2021sw, Papp.2021gaj}, and differs from temporal multiplexing where virtual nodes are constructed with delayed feedback~\cite{Appeltant:2011jy, Paquot:2012dx, Ortin:2015au, Larger:2017cx, Zong:2021dm, Watt:2021ia}. The latter also includes reservoirs based on optical cavities where multimode dynamics (such as frequency combs) are exploited but the output spaces are still constructed by temporal multiplexing \cite{butschek2022photonic,vatin2018enhanced,harkhoe2019delay}.

We illustrate the concept of modal multiplexing with a pattern recognition task using a magnon-scattering reservoir (MSR). The patterns comprise a sequence of symbols "A" and "B" represented by radiofrequency (rf) signals, which consist of sine wave pulses with two distinct frequencies, $f_A$ and $f_B$, and amplitudes $b_{\mathrm{rf},A}$ and $b_{\mathrm{rf},b}$ as shown in Fig.~\ref{fig:FIG1}D. An example of the power spectrum of the input sequence is given in Fig.~\ref{fig:FIG1}E. The rf pulses generate oscillating magnetic fields along the $z$ direction through an $\Omega$-shaped antenna, which surrounds a 5\,$\mu$m wide, 50\,nm thick $\mathrm{Ni}_{81}\mathrm{Fe}_{19}$ disk which hosts a magnetic vortex as a ground state (Fig.~\ref{fig:FIG1}F). $f_A$ and $f_B$ are chosen to coincide with the frequencies of radial eigenmodes of the vortex, which, when excited above a given threshold, result in the excitation of azimuthal eigenmodes through three-magnon scattering processes~\cite{schultheissExcitationWhisperingGallery2019}. The power spectrum of excited magnons is obtained through micro-focused Brillouin light scattering spectroscopy ($\mu$BLS), where a portion of the disk is probed (see supplementary materials). In the linear response regime, we note that neither the input spectrum (Fig.~\ref{fig:FIG1}E) nor the directly-excited magnon spectrum (Fig.~\ref{fig:FIG1}G) gives any information about the actual sequence of "A" and "B" (e.g., "AB" and "BA" are equivalent). When nonlinear processes are at play, however, magnon scattering and associated transient processes result in distinct spectral signatures that can be used to distinguish between different input sequences (Fig.~\ref{fig:FIG1}H).

\begin{figure*}[h!]
    \centering
    \includegraphics[width=\textwidth]{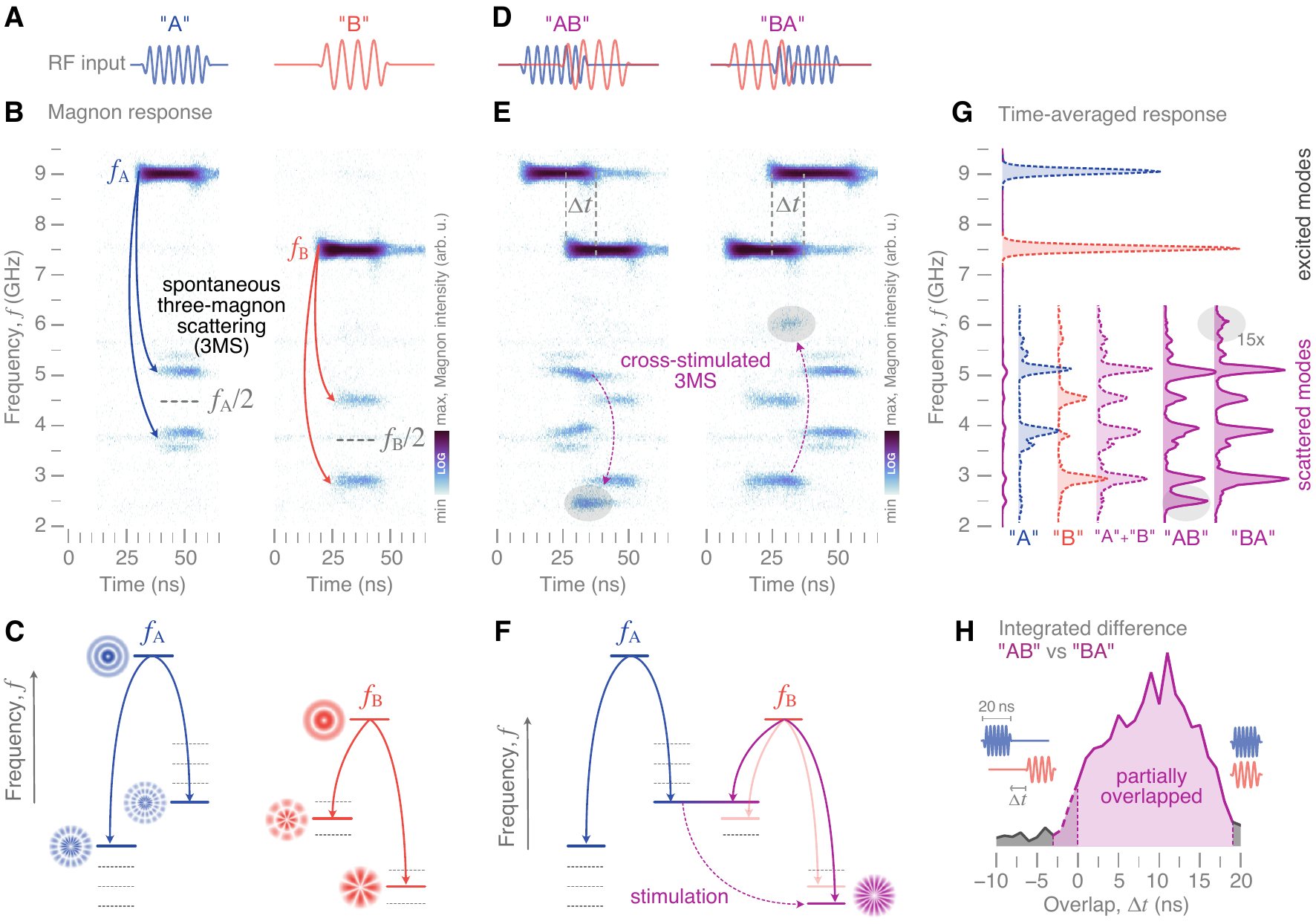}
    \caption{\textbf{Physical background of the magnon-scattering reservoir (MSR).} When pumped strongly by microwave fields (\textbf{A}), a directly-excited primary magnon splits into two secondary magnons (\textbf{B}) via spontaneous 3MS (\textbf{C}). (B) shows the time-resolved frequency response of the MSR to two different input frequencies measured with TR-$\mu$BLS. (\textbf{D}) Driving the MSR with two different, temporally overlapping microwave pulses "A" and "B" leads to (E-F) cross-stimulated 3MS between the channels and to additional peaks in the frequency response. (\textbf{G}) Shows the time-averaged output spectra which are different depending on the temporal order of the pulses. (\textbf{H}) The integrated difference between the spectra of "AB" and "BA" shows that the responses are different when the pulses overlap in time.}
    \label{fig:FIG2}
\end{figure*}

Figure~\ref{fig:FIG2} illustrates the role of three-magnon scattering (3MS), the primary nonlinear process at play for the MSR, in which a strongly-excited primary magnon splits into two secondary magnons under conservation of energy and momentum. We choose 20-ns pulses of $f_\mathrm{A}=8.9$\,GHz (20\,dBm) and $f_\mathrm{B}=7.4$\,GHz (24\,dBm), which excite different radial modes of the vortex, to represent the symbols "A" and "B", respectively (Fig.~\ref{fig:FIG2}A). The magnon intensity is probed as a function of frequency and time using TR-$\mu$BLS (see supplementary materials) and is color-coded in Fig.~\ref{fig:FIG2}B. We measure not only the directly excited primary magnons at $f_\mathrm{A}$ and $f_\mathrm{B}$, but also magnons at frequencies around half the respective excitation frequencies which result from the nonlinearity of spontaneous 3MS (see Fig.~\ref{fig:FIG2}C) \cite{schultheissExcitationWhisperingGallery2019, korberNonlocalStimulationThreemagnon2020}. Here, only the scattering channel with the lowest power threshold is active while other allowed scattering channels remain silent (depicted by dotted lines in Fig.~\ref{fig:FIG2}C).

Cross-stimulation occurs when signals "A" and "B" overlap in time, as shown in Fig.~\ref{fig:FIG2}D. Two different primary magnons that share a common secondary mode, as is depicted in Fig.~\ref{fig:FIG2}F, can result in two 3MS channels that mutually cross-stimulate each other, even below their intrinsic thresholds and along silent channels~\cite{korberNonlocalStimulationThreemagnon2020}. Thus, within the overlap interval, the pumped secondary magnon of the first symbol influences the primary mode scattering of the second symbol, and vice versa, leading to the primary mode scattering into multiple pairs of secondary modes (Fig.~\ref{fig:FIG2}E).

Because cross-stimulation strongly depends on the temporal order of the primary excitation, it provides an important physical resource for processing the temporal sequence of our "AB" signals. This is shown in Fig.~\ref{fig:FIG2}G, where we compare the time-averaged power spectra for the "AB" and "BA" sequences. These spectra are computed from integrating the temporal data in Fig.~\ref{fig:FIG2}E. When only signal "A" or only signal "B" is applied, we measure conventional spontaneous 3MS of the respective primary modes with the secondary modes already discussed above in context of Fig.~\ref{fig:FIG2}B. Within the overlap interval, the mutual cross-stimulation leads to additional peaks in the scattered mode spectrum. As highlighted by shaded areas in Fig.~\ref{fig:FIG2}E, the frequencies of these additional scattered modes strongly depend on the temporal order of the two input signals. Consequently, the average spectra of "AB" and "BA" are different from each other, and neither can be constructed from a simple superposition of the average spectra of "A" and "B" individually (Fig.~\ref{fig:FIG2}G). This is the key principle that underpins how the MSR processes temporal signals.

To highlight the significance of the transient times, we vary the overlap $\Delta t$ of the symbols "A" and "B" and determine the frequency-averaged difference between the time-averaged spectra of "AB" and "BA" (Fig.~\ref{fig:FIG2}H). This difference is zero when the two input pulses do not overlap since no cross-stimulation takes place. With increasing overlap, however, cross-stimulation between the two pulses becomes more significant and leads to a difference in the output of the reservoir. This difference vanishes again when the input pulses fully overlap and, thus, arrive at the same time.

\begin{figure}[h!]
    \centering
    \includegraphics{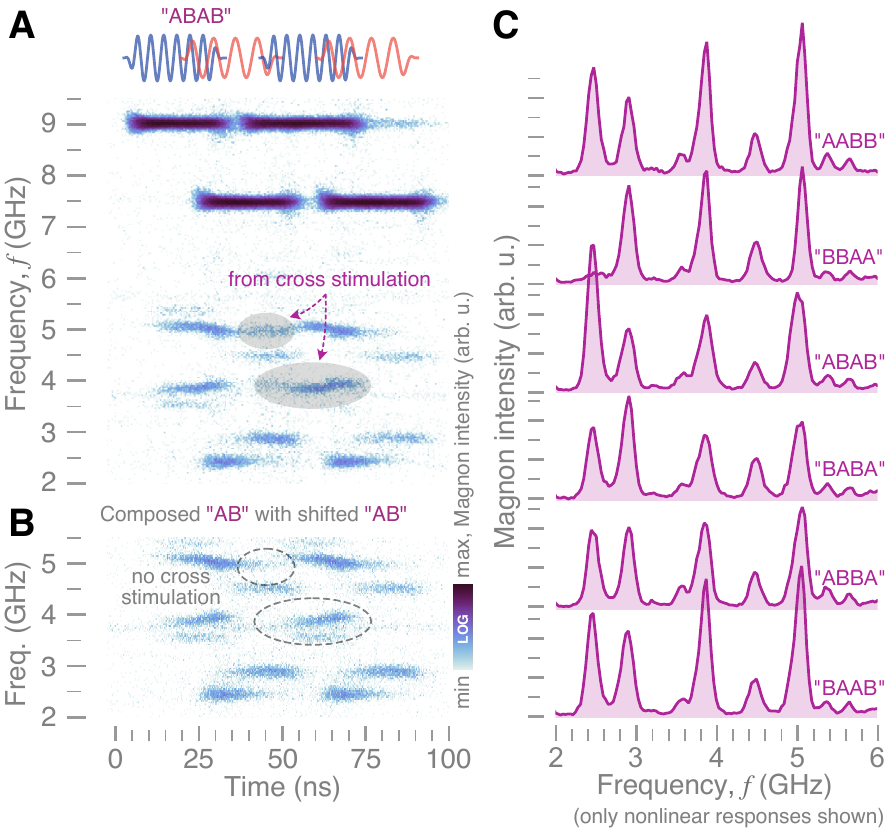}
    \caption{\textbf{Performance of the MSR for longer temporal patterns.} (\textbf{A}) Time-resolved spectral response of the MSR to a four-symbol microwave pattern "ABAB", detected with TR-$\mu$BLS. (\textbf{B}) For reference, the spectrum of "AB" is overlayed with a shifted version of itself. Differences between composed and real spectrum (due to cross-stimulated magnon scattering) are highlighted by shaded and circled areas. (\textbf{C}) Average output spectra of the MSR for different four-symbol patterns with the same average input-frequency content but clearly different nonlinear response.}
    \label{fig:FIG3}
\end{figure}

In order to explore the capabilities of the presented MSR, the complexity of the input signals was further increased. Figure~\ref{fig:FIG3}A shows the nonlinear response to the four-symbol pulse pattern "ABAB". In contrast to a reference spectrum composed by a simple linear superposition of two consecutive "AB" patterns, shown in Fig.~\ref{fig:FIG3}B, the real spectrum of the four-symbol response contains additional features which are generated by cross-stimulated scattering when two pulses overlap. The differences are highlighted by the shaded areas in Fig.~\ref{fig:FIG3}A and circled areas in Fig.~\ref{fig:FIG3}B, respectively. This behavior illustrates that cross-stimulation can result in distinct features that allow to distinguish also longer patterns. This is further exemplified in Fig.~\ref{fig:FIG3}C, which shows  the six four-symbol combinations comprising two "A" and two "B". Like the data in Fig.~\ref{fig:FIG2}, transient processes from cross-stimulation generate distinct power spectra for the six combinations, which would be indistinguishable in the linear response regime.

\begin{figure}[h!]
    \centering
    \includegraphics{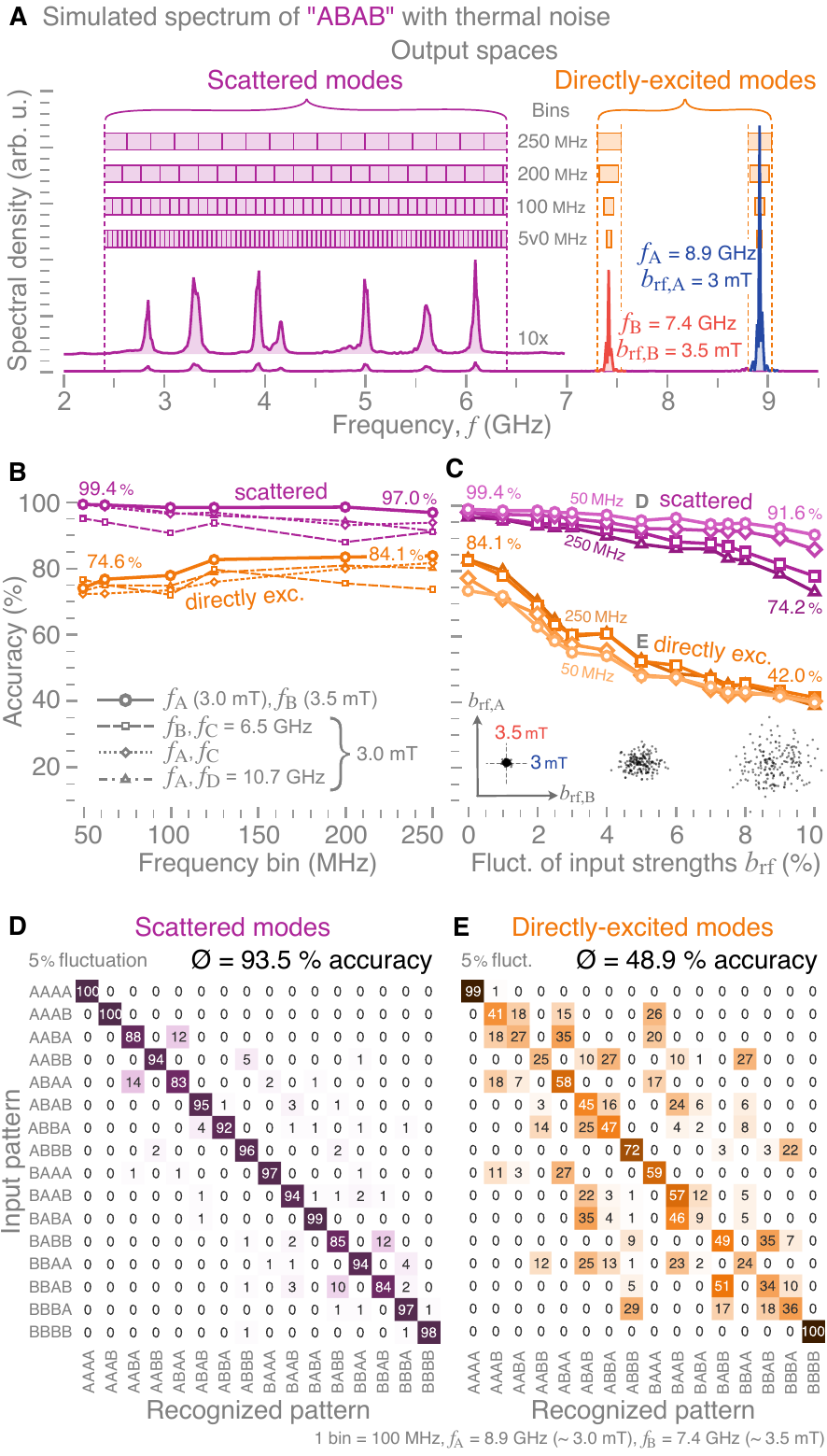}
    \caption{\textbf{Micromagnetic modeling of pattern recognition capabilities.} (\textbf{A}) Simulated spectrum of the pattern "ABAB" with the definition of different output spaces (scattered and directly excited modes) for the MSR. (\textbf{B}) Average detection accuracy of four-symbol patterns for different output spaces and excitation frequency and  power combinations as a function of frequency bin sizes. (\textbf{C}) Accuracy for different output spaces and bin sizes as a function of power fluctuations in the input signals (depicted by the insets). (\textbf{D}), (\textbf{E}) Corresponding confusion matrices for the two output spaces, respectively, both for a the same frequency combination, bin size and input power fluctuation.}
    \label{fig:FIG4}
\end{figure}

Since the experimental data requires the integration of thousands pulse cycles, we rely on micromagnetic simulations to quantify the capacity of the MSR for recognizing all possible combinations of four-symbol sequences composed from "A" and "B". Thereby, we are able to analyze individual pulse sequences and study the influence of thermal noise and amplitude fluctuations on the recognition rate of the MSR.
Figure~\ref{fig:FIG4}A shows a simulated power spectrum (at $T = 300$ K) for the input pattern "ABAB" with $f_\mathrm{A}=8.9$\,GHz ($b_\mathrm{rf,A}=3$\, mT) and $f_\mathrm{B}=7.4$\,GHz ($b_\mathrm{rf,B}=3.5$\,mT). The output spaces of the reservoir are defined by subdividing the time-averaged power spectrum into frequency bins of different widths. To emphasize the importance of the scattering (interconnection) between the different magnon modes, we study the performance of the MSR for two separate output spaces (Fig.~\ref{fig:FIG4}A). One output space for the scattered modes is constructed over a 4-GHz window below $f_A$ and $f_B$, where the different frequency bins result in an output vector with 16 to 80 components depending on the bin size (see supplementary materials). For comparison, a two-dimensional output space corresponding to the directly-excited modes is constructed by averaging within bins centered around $f_A$ and $f_B$.

For each four-symbol sequence, 200 simulations were executed with different realizations of the thermal field in order to generate distinct output states. Supervised learning using logistic regression was then performed on this data set to construct trained models of the output states based on either the directly-excited or scattered modes. The accuracy of these models for different combinations of input frequencies $f_\mathrm{A}=8.9$\,GHz, $f_\mathrm{B}=7.2$\,GHz, $f_\mathrm{C}=6.5$\,GHz, $f_\mathrm{D}=10.7$\,GHz (and corresponding input strengths $b_{\mathrm{rf},i}$) is shown in Fig.~\ref{fig:FIG4}B as a function of bin size. We find that the MSR performs comparably well when choosing different input frequencies (different radial modes) to represent the input symbols. Hence, an extension of the input space to more than two frequencies/symbols ("A","B","C", "D", etc.), or even to more broadband signals, is straightforward.

Overall, the accuracy depends weakly on the bin size. The recognition rate slightly increases with increasing bin size for the directly-exited modes whereas it decreases marginally for the scattered modes. This can be understood from the fact that smaller bin sizes capture more features of the power spectrum of the scattered modes, while for the directly-excited modes the larger bin sizes contain more information about potential nonlinear frequency shifts, which helps to separate the inputs. We observe that outputs based on the directly-excited modes can yield an accuracy of $\sim$84\%, while scattered modes provide a notable improvement in performance, with an accuracy reaching 99.4\% for the case considered in Fig.~\ref{fig:FIG4}A.

In general, the scattered modes provide higher accuracy for pattern recognition compared with the directly-excited modes. The difference in accuracy becomes even more pronounced when amplitude fluctuations are present.
Figure~\ref{fig:FIG4}C illustrates how the accuracy evolves with the fluctuation strength, which represents the width of the normal distributions (in \%), centered around the nominal values of $b_\mathrm{rf,A}$ and $b_\mathrm{rf,B}$, from which the field strengths are drawn, as shown in the inset for $b_\mathrm{rf,A}=3$\,mT and $b_\mathrm{rf,B}=3.5$\,mT.
The performance of the MSR based on the directly-excited modes drops significantly with increasing fluctuation strength to (42\% accuracy at 10\% fluctuation). However, recognition based on the scattered modes is much more resilient, with a decrease to only between $\sim$74\% and $\sim$92\% accuracy (depending on the bin size).

Figures~\ref{fig:FIG4}E,F show confusion matrices for the scattered and directly-excited modes, respectively, both for the same set of parameters. They highlight the robustness of the MSR which is based on the scattered modes since it mainly fails to distinguish "AABA" from "ABAA" and "BBAB" from "BBAB" in $\sim$12\% of the cases.
The MSR based on the directly-excited modes, on the other hand, fails to recognize almost all of the patterns, except for the trivial cases of "AAAA" and "BBBB" for which there is practically no ambiguity in the inputs. These trends do not depend on the type of supervised learning used and highlight the important role of cross-stimulated 3MS in the MSR for the pattern recognition of noisy radiofrequency signals.

Our findings demonstrate the possibility of performing reservoir computing in modal space utilizing the intrinsic nonlinear properties of a magnetic system, namely the scattering processes between magnons in a magnetic vortex disk. Temporal patterns encoded using two different input-frequency pulses can be distinguished with high accuracy. The results also indicate that input patterns can be extended to more broadband signals. We note that the technical design of the physical reservoir is extremely simple and requires very little prepossessing, while the complexity of the data handling arises mostly from the intrinsic nonlinear dynamics of the magnon system.

\section*{Author declarations}

\subsection*{Conflict of Interest}
The authors have no conflicts to disclose.

\subsection*{Author's contributions}

Conceptualization: HS, JVK, KS. Investigation: CH. Simulation: JVK. Visualization: LK, TH, JVK, HS. Funding acquisition: JVK, HS, JF. Project administration: HS, KS. Writing – original draft: LK. Writing – review and editing: LK, CH, TH, JVK, HS, JF, KS.

\section*{Acknowledgements}

The authors are thankful to D. Rontani and K. Knobloch for providing feedback on the manuscript and fruitful discussions. This study was supported by the German Research Foundation (DFG) within programs SCHU 2922/1-1, KA 5069/1-1 and KA 5069/3-1, as well as by the French Research Agency (ANR) under contract No. ANR-20-CE24-0012 (MARIN). Support by the Nanofabrication Facilities Rossendorf (NanoFaRo) at the IBC is gratefully acknowledged.

\section*{Data availability}
The data that support the findings of this study will be made openly available at RODARE.

%\section*{References}

%\bibliography{references.bib,new_refs_jvk.bib}

\end{document}